\documentclass[journal]{IEEEtran}

\usepackage{amsmath,amssymb}
\usepackage{booktabs,tabularx,array,multirow}
\usepackage{graphicx}
\usepackage{xcolor}
\usepackage{url}
\usepackage{cite}
\usepackage{tikz}
\usetikzlibrary{arrows.meta,positioning,fit,calc,shapes.geometric,backgrounds}
\usepackage{enumitem}
\usepackage{balance}

\newcommand{\mestar}{ME-STAR-RIS}

\title{\huge Movable-Element STAR-RIS for 6G Non-Terrestrial Networks: Architectures, Enabling Technologies, and Open Challenges}

\author{Wali Ullah Khan,~\IEEEmembership{Member,~IEEE} and Muhammad Adil
\thanks{Wali Ullah Khan is with SnT, University of Luxembourg, Luxembourg (e-mail: waliullahkhan30@gmail.com).}
\thanks{Muhammad Adil is with the Department of Electronics Engineering, University of Rome Tor Vergata, 00133 Rome, Italy (e-mail: adil34700@gmail.com).}
}

\begin{document}
\maketitle

\begin{abstract}
Non-terrestrial networks (NTNs) are emerging as a key component of
6G systems by extending ubiquitous and resilient connectivity across
satellite, aerial, and terrestrial layers. However, severe propagation
loss, rapidly varying geometry, Doppler, blockage, and stringent
payload and energy constraints make efficient beam and propagation
control particularly challenging. Movable-element STAR-RIS
(ME-STAR-RIS) offers a new approach by combining full-space
transmission/reflection control with reconfigurable surface geometry,
thereby introducing an additional spatial degree of freedom for
geometry-aware NTN operation. In this article, we first describe the operating principle, movement constraints, and potential hardware realizations of ME-STAR-RIS, and then present representative satellite-borne, aerial-relay, and
cross-tier NTN architectures. We discuss the key enabling mechanisms,
including joint geometry--electromagnetic optimization,
mobility-aware channel acquisition, multi-timescale control, and
predictive configuration. A LEO-centered numerical case study
compares ME-STAR-RIS with an otherwise identical fixed STAR-RIS and
demonstrates the benefit of element mobility across different transmit
powers, movement ranges, and surface sizes, while revealing
diminishing returns from increasingly large displacements. Finally,
we discuss the main hardware, CSI, scalability, near-field, sensing,
security, reliability, and standardization challenges, and highlight
promising directions toward practical ME-STAR-RIS-enabled NTNs.
\end{abstract}

%\begin{IEEEkeywords}Movable-element STAR-RIS, ME-STAR-RIS, simultaneously transmitting and reflecting reconfigurable intelligent surface, non-terrestrial networks, satellite communications, 6G, geometry-aware beamforming.\end{IEEEkeywords}

\section{Introduction}
\IEEEPARstart{N}{on-terrestrial} networks (NTNs) are evolving from specialized satellite systems toward an integral component of beyond-5G and 6G connectivity \cite{11674230}. Their role spans broadband access, direct-to-device services, emergency communications, Internet-of-Things connectivity, positioning, sensing, and integrated space--air--ground services. The resulting network is highly heterogeneous: low-Earth-orbit (LEO) satellites move rapidly relative to users; high-altitude platforms (HAPs) and unmanned aerial vehicles (UAVs) introduce additional mobility; and terrestrial gateways must coordinate with multiple orbital and aerial nodes. These characteristics create coverage and capacity opportunities, but also amplify propagation loss, Doppler, blockage, handover, synchronization, and payload-energy challenges \cite{10716670}.

Reconfigurable intelligent surfaces (RISs) have been widely considered for NTN link enhancement because they can reshape propagation with relatively low RF-chain complexity. Recent satellite-centric studies have emphasized that the usefulness of RISs in space systems depends strongly on deployment geometry, visibility, Doppler, control latency, and payload constraints rather than on passive beamforming alone \cite{10716670,zheng2026satris}. Yet conventional reflecting RISs predominantly serve one half-space. STAR-RISs overcome this limitation by simultaneously transmitting and reflecting the impinging signal, thereby supporting users on both sides of the surface. Energy-splitting, mode-switching, and time-switching protocols provide different realizations of this full-space control \cite{10133841,liu2022coupled}. Their full-space benefit has also been demonstrated in UAV-assisted air-to-ground communications, where the platform trajectory and STAR coefficients are jointly adapted while the surface geometry remains fixed \cite{zhao2022staruav}.

\begin{figure*}[t]
\centering
\includegraphics[width=0.92\textwidth]{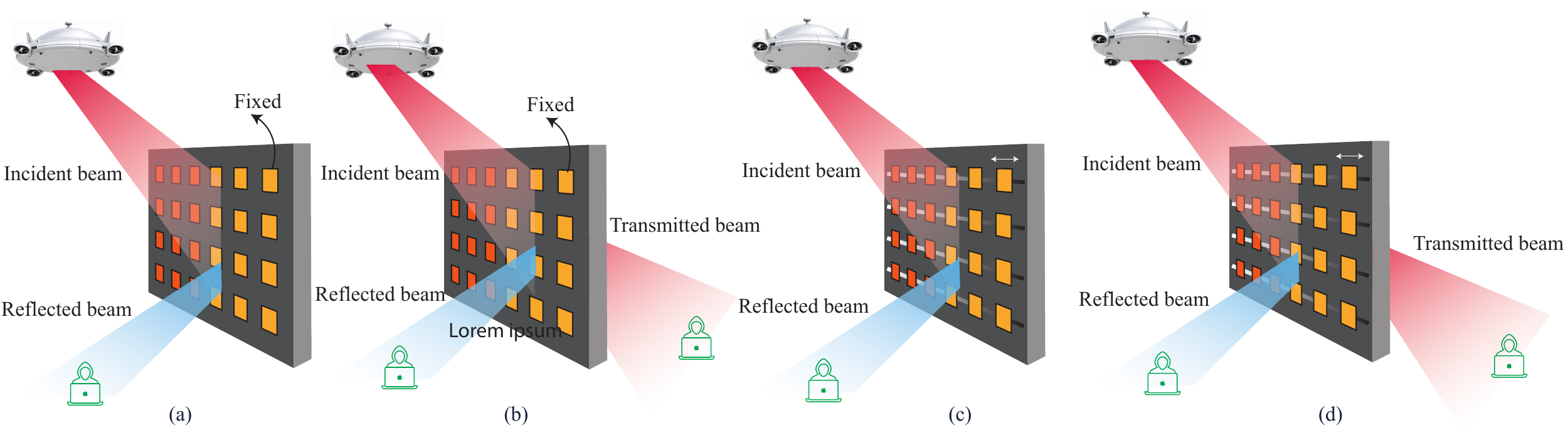}
\caption{Evolution from a fixed reflecting RIS to a
geometry-reconfigurable ME-STAR-RIS: (a) conventional RIS,
(b) STAR-RIS, (c) ME-RIS, and (d) ME-STAR-RIS. The key additional
degree of freedom is the ability to alter the physical element
locations while retaining STAR control.}
\label{fig:evolution}
\end{figure*}

A remaining limitation is less visible but fundamental: the physical coordinates of the surface elements are normally fixed. Recent movable-antenna and movable-element RIS research has shown that local position reconfiguration can exploit spatial channel variation and create additional geometric degrees of freedom \cite{zhu2024ma,zhu2025tutorial,zhou2026meris,zhao2026meris}. More recently, movable-element simultaneously transmitting and reflecting surfaces have been investigated for terrestrial communication, security, and integrated sensing scenarios, demonstrating the value of jointly optimizing element locations and electromagnetic coefficients \cite{zhao2026mestars,zhao2026secure,zhou2025iscc}. A related study combining movable antennas with STAR-RISs at the transceiver side further reinforces the broader value of joint spatial and electromagnetic adaptation \cite{amiri2025swipt}. However, these emerging ME-STARS and movable-antenna/STAR-RIS studies are predominantly terrestrial, while satellite-RIS and STAR-RIS aerial studies still assume fixed element locations. This leaves an NTN-level gap: how should full-space STAR control and local aperture-geometry reconfiguration be combined under orbital mobility, Doppler, multi-tier deployment, and actuation constraints?

\begin{table*}[t]
\caption{Comparison of RIS, STAR-RIS, ME-RIS, and ME-STAR-RIS}
\label{tab:compare}
\centering
\footnotesize
\begin{tabularx}{\textwidth}{|p{2.4cm}|X|X|X|X|}
\hline
\textbf{Feature} & \textbf{RIS} & \textbf{STAR-RIS} & \textbf{ME-RIS} & \textbf{ME-STAR-RIS} \\
\hline\hline
Coverage domain & Mainly reflection half-space & Transmission and reflection full-space & Mainly reflection half-space & Transmission and reflection full-space \\ \hline
Element coordinates & Fixed & Fixed & Reconfigurable & Reconfigurable \\ \hline
Primary control variables & Reflection phase/amplitude & T/R phase and amplitude; protocol state & Reflection coefficients and element positions & T/R coefficients, element positions and protocol state \\ \hline
Spatial DoF & Fixed by physical aperture geometry & Fixed by physical aperture geometry & Geometry-adaptive & Geometry-adaptive on both sides of the surface \\ \hline
Potential NTN role & Link redirection, blockage bypass & Full-space coverage and user grouping & Geometry-aware reflection and interference shaping & Joint aperture shaping, full-space coverage, and geometry-aware interference control \\ \hline
Main added cost & Control circuitry & T/R hardware and control coupling & Position actuation and calibration & T/R hardware, actuation and position-aware calibration \\
\hline
\end{tabularx}
\end{table*}

This motivates a broader question: \emph{what if an NTN could reconfigure not only the electromagnetic state of a STAR-RIS, but also its aperture geometry?} Such a surface could adapt its array response to changing satellite-user geometry, reshape spatial correlation, create or suppress path diversity, and reallocate its effective aperture between transmission and reflection regions. In this article, we use the term \mestar{} for this concept; the closely related term \emph{movable-element STARS (ME-STARS)} is also used in the recent literature \cite{zhao2026mestars}.

The main contributions of this article are fourfold. First, we present an NTN-oriented interpretation of ME-STAR-RIS and clarify how geometry reconfiguration complements conventional STAR coefficient control. Second, we develop three representative deployment architectures spanning satellite, aerial, and terrestrial network tiers. Third, we identify the mechanisms required to operate ME-STAR-RIS under high mobility, including position-dependent channel acquisition, Doppler-aware control, and multi-timescale optimization. Fourth, we expose the practical tradeoffs introduced by physical movement and validate the concept through a LEO-centered numerical case study that compares ME-STAR-RIS directly with an otherwise identical fixed STAR-RIS.

\section{From Static STAR-RIS to Geometry-Reconfigurable Surfaces}
\subsection{Evolution of the Programmable Aperture}
Fig.~\ref{fig:evolution} summarizes the conceptual evolution. A conventional RIS controls reflection coefficients over a fixed aperture. A STAR-RIS adds simultaneous transmission and reflection, enabling full-space coverage but retaining a fixed geometry. A movable-element RIS (ME-RIS) introduces position control but remains reflection-oriented. ME-STAR-RIS combines both capabilities: each element (or a small element group) can change its position within a feasible region while also controlling the transmitted and reflected fields. Table~\ref{tab:compare} contrasts the four surface types from an NTN perspective. The essential distinction is that ME-STAR-RIS creates a geometry--electromagnetic design space rather than an electromagnetic-only design space.

\subsection{Operating Principle and Movement Constraints}
A magazine-level description is sufficient to capture the additional degree of freedom introduced by movable elements. Each ME-STAR-RIS element is assigned a feasible local movement region, such as a short one-dimensional track, a small two-dimensional cell, or a group-wise displacement region. The controller selects the element positions while preserving a minimum inter-element spacing so that collisions and excessively strong mutual coupling are avoided. The selected geometry directly changes the phase progression, spatial correlation, and effective array response seen by the incident and outgoing waves.

\begin{table*}[t]
\caption{Representative ME-STAR-RIS NTN Architectures: Roles, Control Timescales, and Practical Limits}
\label{tab:deploy}
\centering
\footnotesize
\begin{tabularx}{\textwidth}{|p{1.6cm}|p{2.8cm}|p{3.0cm}|p{2.8cm}|p{3.0cm}|X|}
\hline
\textbf{Architecture} & \textbf{Primary role} & \textbf{Why movement helps} & \textbf{Control hierarchy} & \textbf{Main coupling} & \textbf{Dominant practical concern} \\
\hline\hline
Satellite-borne & Multi-beam access, interference shaping, full-space payload/relay functions & Adapts the effective aperture to recurring orbital and user geometry & Slow or event-triggered geometry; fast STAR coefficients & Orbit geometry, element positions, STAR coefficients and satellite beamforming & Space qualification, vibration, thermal stability, actuator lifetime and power \\ \hline
Aerial-relay & Blockage bypass, hotspot coverage, temporary NTN relay & Adds local aperture adaptation on top of UAV/HAPS platform mobility & Slow platform motion; medium element repositioning; fast electronic control & Platform trajectory, element positions and STAR coefficients & Propulsion--actuation energy tradeoff and multi-timescale control \\\hline
Cross-tier integrated & Gateway enhancement, NTN/TN handover, multi-connectivity, cell-edge support & Reshapes local scattering and spatial correlation without the strictest spacecraft constraints & Medium geometry updates; fast STAR/beamforming updates & Satellite association, terrestrial resources, geometry and STAR coefficients & Signaling, synchronization, calibration, and multi-domain coordination \\
\hline
\end{tabularx}
\end{table*}

The electronic STAR operation is then configured on top of this geometry. Under a commonly used energy-splitting interpretation, each element divides the incident power between transmission and reflection while controlling the corresponding phases. For exposition, the transmitted and reflected power fractions may be viewed as nonnegative quantities whose sum is one. This ideal description is useful for architecture-level discussion, but practical passive STAR hardware can impose coupling between the transmission and reflection phases and amplitudes \cite{liu2022coupled}. Hence, the feasible electromagnetic response can depend on the hardware state and, in a movable implementation, potentially on the selected element geometry as well.

The important point is that the cascaded NTN channel now depends on two coupled control layers: the physical element locations and the STAR transmission/reflection states. Moving an element changes the channel itself; electronic STAR control then operates on that geometry-dependent channel. For NTNs, the movement region cannot be treated as an unconstrained design space. The allowed displacement must account for payload mass, structural stiffness, thermal conditions, launch vibration, actuator lifetime, calibration accuracy, and control wiring. Consequently, limited-stroke or group-wise motion is likely to be more realistic than frequent arbitrary element-by-element movement, particularly for space-borne platforms.

\subsection{Candidate Hardware Realizations}
Movable-antenna research provides a useful implementation starting point, including motorized positioning, mechanical tracks, MEMS-scale actuation, and electronically reconfigurable structures that emulate spatial repositioning \cite{zhu2024ma,zhu2025tutorial}. Translating these concepts to STAR-RIS requires additional care because each moving element must preserve both its transmission and reflection response while maintaining stable biasing/control connections. For terrestrial gateways and UAV payloads, electromechanical tracks may be plausible for relatively slow repositioning. For spacecraft, reliability and radiation tolerance become dominant; limited-stroke, group-wise, or quasi-static movement may be more realistic than frequent per-element motion. These hardware pathways should therefore be viewed as research candidates rather than established space-qualified solutions.

\section{ME-STAR-RIS-Empowered NTN Architectures}

%\section{ME-STAR-RIS-Empowered NTN Architectures}

\begin{figure}[t]
\centering
\includegraphics[width=\columnwidth]{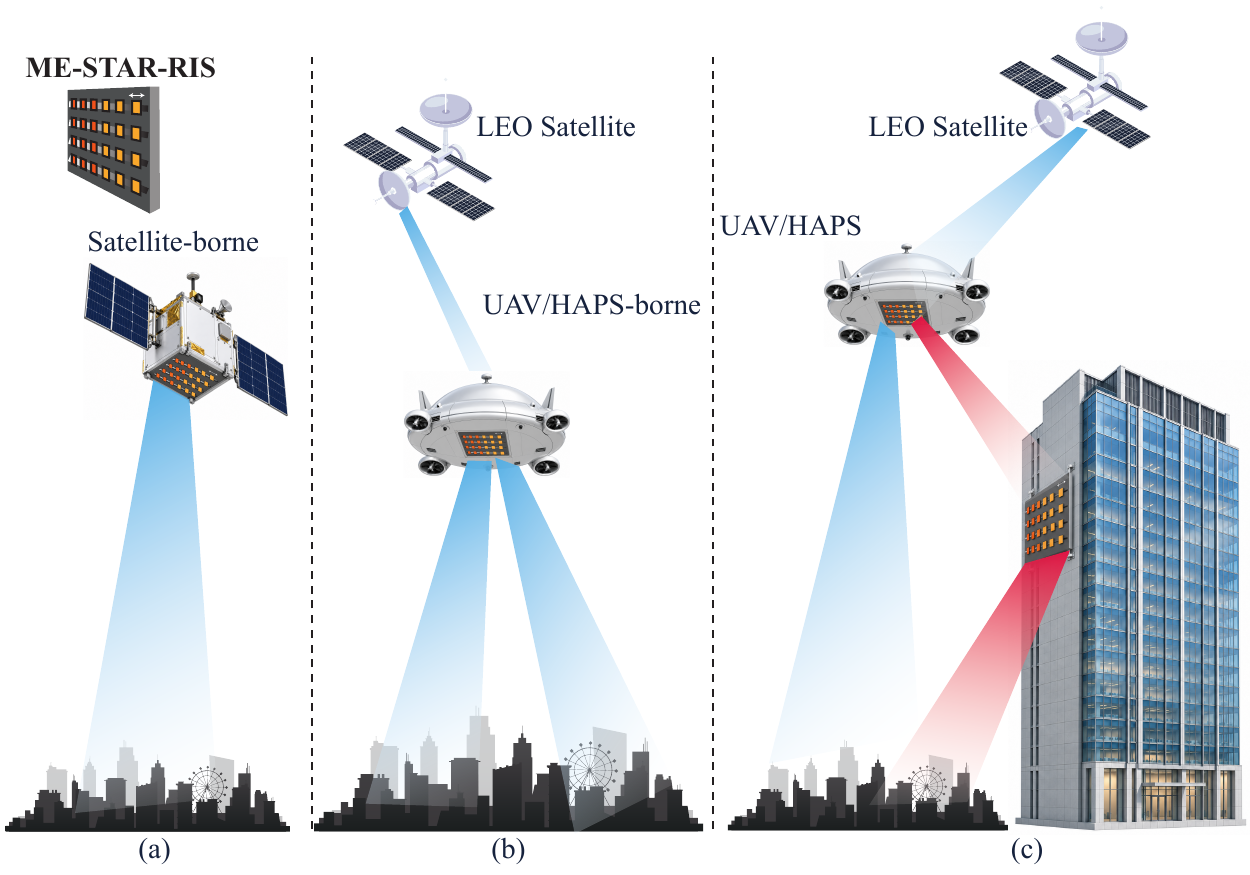}
\caption{Representative ME-STAR-RIS deployment architectures for NTNs:
(a) Satellite-borne ME-STAR-RIS, (b) Aerial-relay ME-STAR-RIS, and (c) Cross-tier integrated ME-STAR-RIS involving
non-terrestrial and terrestrial infrastructure.}
\label{fig:architectures}
\end{figure}

The value of movable elements depends strongly on where the surface is
deployed. Fig.~\ref{fig:architectures} illustrates three representative
architectures spanning the space, air, and terrestrial layers, namely
satellite-borne, aerial-relay, and cross-tier integrated ME-STAR-RIS.
Table~\ref{tab:deploy} complements this illustration by comparing
their primary roles, benefits of element movement, control hierarchies,
coupled design variables, and dominant practical limitations.

\subsection{Satellite-Borne ME-STAR-RIS}
A satellite-borne ME-STAR-RIS can be integrated with a payload antenna or a conformal external panel. In this architecture, the surface becomes part of the satellite beamforming front end. Conventional STAR coefficients provide fast phase/amplitude adaptation, whereas limited element repositioning reshapes the effective aperture more slowly. This is particularly attractive when a fixed regular array produces persistent correlation, grating-lobe, or interference patterns for a recurring service geometry.

The architecture may support full-space functions that a purely reflecting payload cannot easily provide. Depending on how the panel is installed and illuminated, one side of a properly oriented STAR aperture could support an Earth-facing access or feeder function, while the opposite side supports a space-facing relay or inter-satellite function. Element movement would then reshape the effective aperture more slowly according to recurring orbital geometry, traffic distribution, elevation angle, or interference conditions, while electronic STAR coefficients provide faster adaptation. A statistical-CSI philosophy may be especially important because obtaining instantaneous cascaded CSI in LEO is difficult; related work on satellite-mounted programmable metasurfaces has already highlighted the role of statistical CSI and low-complexity wave-domain processing \cite{lin2024simleo}.

The principal limitation is hardware realism. Frequent mechanical movement aboard a satellite can consume energy, introduce vibration, complicate thermal design, and reduce reliability. Consequently, satellite-borne ME-STAR-RIS is best viewed as a \emph{slow geometry reconfiguration} mechanism complemented by much faster electronic STAR updates.

\subsection{Aerial-Relay ME-STAR-RIS}
The second architecture mounts ME-STAR-RIS on a UAV or HAPS. This creates two layers of spatial reconfigurability: the platform changes its global position while the surface elements change their local geometry. Existing STAR-RIS-assisted UAV work already shows the value of coordinating aerial mobility with full-space transmission/reflection control \cite{zhao2022staruav}; ME-STAR-RIS adds local aperture-geometry adaptation as a further degree of freedom. The platform can first move toward a favorable macro-location, after which element positioning refines the local aperture response and STAR coefficients manage the transmitted/reflected fields.

This architecture is attractive for temporary coverage, disaster recovery, hotspot offloading, and satellite-to-ground blockage bypass. It also exposes a new control problem: platform motion, element motion, and STAR coefficient adaptation operate at different timescales and have different energy costs. A naive controller that continuously moves both the UAV and surface elements may waste energy. Instead, the platform trajectory should respond to large-scale geometry while element movement corrects residual channel structure, interference, or user clustering.

\subsection{Cross-Tier Integrated ME-STAR-RIS}
In the third architecture, ME-STAR-RIS is deployed at terrestrial gateways, building facades, BSs, HAPS, or selected relay nodes and coordinated with satellites. This approach follows the broader deployment-aware view that the usefulness of RIS-assisted satellite links depends on where controllable apertures are placed in the end-to-end geometry \cite{zheng2026satris}. It avoids the strictest spacecraft actuation constraints while still exploiting geometry reconfiguration where the satellite-ground channel terminates. It can be used to create alternate paths around local blockage, support cell-edge users on both sides of a surface, assist satellite-to-terrestrial handover, or dynamically partition an aperture among NTN and terrestrial services.

Cross-tier operation is also well suited to multi-connectivity. A single surface could simultaneously assist a satellite path and a terrestrial path, with its geometry optimized to reduce correlation between the two services. However, the corresponding optimization spans satellite association, user scheduling, terrestrial beamforming, STAR coefficients, and element locations, making scalable decomposition essential.

\section{Key Enabling Mechanisms}
\subsection{Joint Geometry--Electromagnetic Optimization}
The central ME-STAR-RIS design problem is not simply ``optimize more variables.'' Element motion changes the channel itself; STAR coefficients then operate on that geometry-dependent channel. The natural optimization variables therefore include satellite/BS beamforming, user association, power allocation, element positions, and transmission/reflection coefficients. Recent terrestrial ME-RIS and ME-STARS studies typically use alternating optimization, gradient-based position updates, successive convex approximation, or swarm/heuristic methods to deal with the resulting nonconvexity \cite{zhou2026meris,zhao2026meris,zhao2026mestars}.

For NTN-scale systems, a monolithic optimization is unlikely to meet latency constraints. A more practical strategy is hierarchical. Slow variables, such as satellite selection, surface operating mode, and element geometry, are optimized using ephemeris and long-term channel statistics. Fast variables, such as phase shifts, power, and digital beamforming, are then updated using short-term channel information. This separation is particularly important because the mechanical cost of element motion is fundamentally different from the electronic cost of phase reconfiguration.

\subsection{Mobility- and Doppler-Aware Channel Acquisition}
Position-dependent channels are the enabling resource of ME-STAR-RIS, but they also create a channel-acquisition burden. Estimating the channel only at one surface geometry is insufficient if the elements later move. Brute-force training over all candidate positions is infeasible for large surfaces. Instead, the network should learn or reconstruct a \emph{spatial channel map} over the allowed movement regions, drawing on field-response modeling, compressed measurements, geometric information, and statistical CSI concepts developed for movable antennas and satellite links \cite{zhu2025tutorial,lin2024simleo}.

LEO motion adds another complication: the channel map itself evolves with satellite position and Doppler. The useful object is therefore a time-indexed geometry-aware channel model. Ephemeris information can predict the dominant line-of-sight direction, while pilots and sensing can refine local multipath parameters. This suggests predictive rather than purely reactive ME control.

\begin{figure}[t]
\centering
\includegraphics[width=\columnwidth]{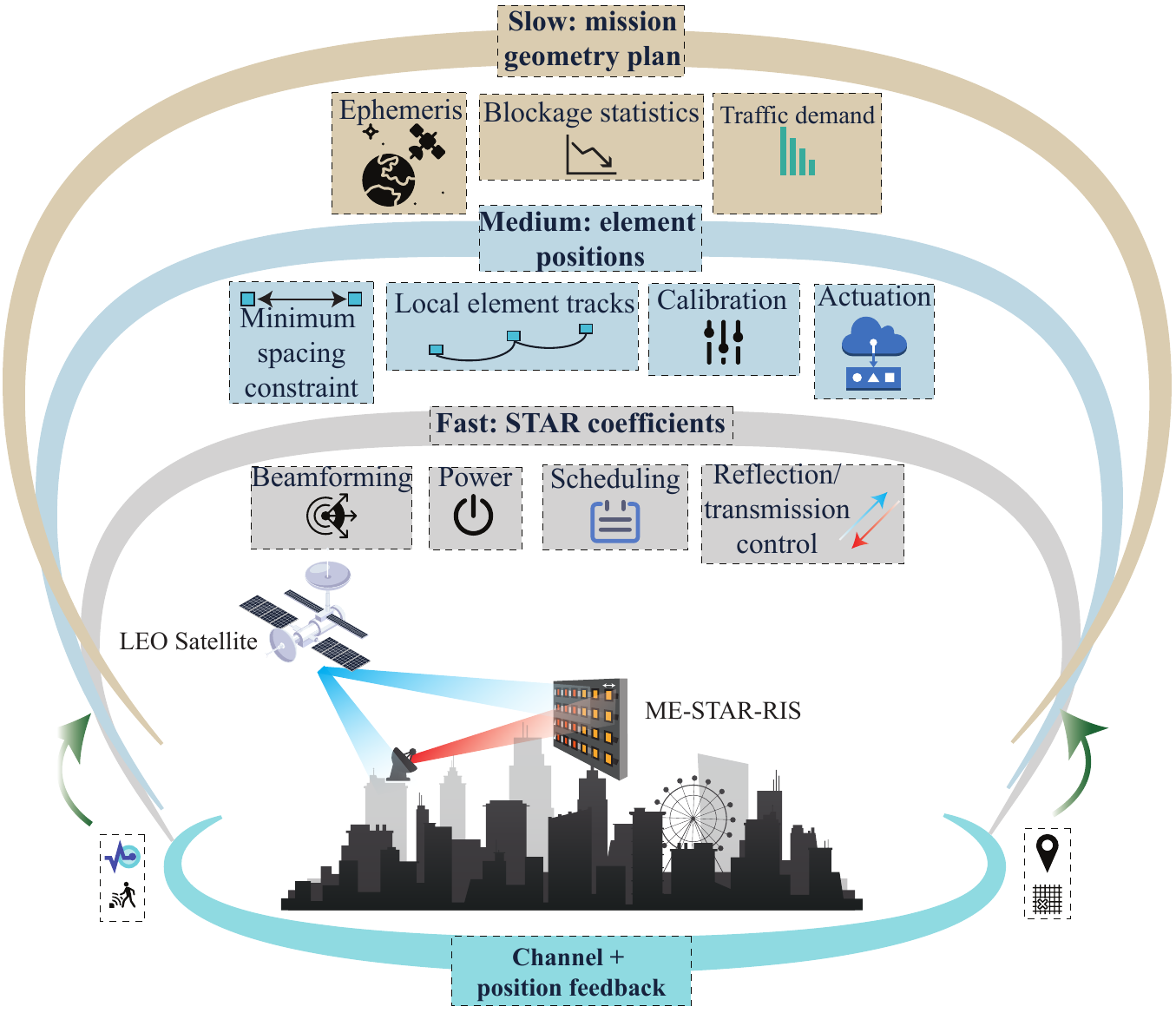}
\caption{Illustrative multi-timescale control framework for
ME-STAR-RIS-enabled NTNs. Long-term orbital geometry, blockage
statistics, and traffic demand guide slow geometry planning; element
positions and actuation are updated at an intermediate timescale; and
electronic STAR coefficients, beamforming, power, and scheduling are
adapted at the fast timescale using channel and position feedback.}
\label{fig:multitimescale}
\end{figure}

\subsection{Multi-Timescale Control}
Fig.~\ref{fig:multitimescale} shows a three-timescale controller. At the slow layer, orbital geometry, service demand, and long-term channel statistics determine whether element movement is needed and which movement region should be activated. At the medium layer, element positions are updated only when the expected gain exceeds an actuation cost threshold. At the fast layer, STAR coefficients and digital beamforming track short-term channel variations without mechanical movement.

% \begin{figure}[t]
% \centering
% \begin{tikzpicture}[font=\scriptsize,>=Latex,node distance=4mm]
% \node[draw,rounded corners,minimum width=7.7cm,minimum height=.72cm] (slow) {\textbf{Slow layer:} ephemeris, traffic, blockage statistics $\rightarrow$ geometry plan};
% \node[draw,rounded corners,below=of slow,minimum width=7.7cm,minimum height=.72cm] (med) {\textbf{Medium layer:} element-position update subject to actuation cost and spacing};
% \node[draw,rounded corners,below=of med,minimum width=7.7cm,minimum height=.72cm] (fast) {\textbf{Fast layer:} STAR coefficients, beamforming, power and scheduling};
% \node[draw,rounded corners,below=of fast,minimum width=7.7cm,minimum height=.72cm] (obs) {Channel/sensing feedback + position/calibration measurements};
% \draw[->,thick] (slow)--(med); \draw[->,thick] (med)--(fast); \draw[->,thick] (fast)--(obs);
% \draw[->,thick] (obs.east) -- ++(.35,0) |- (slow.east);
% \end{tikzpicture}
% \caption{Illustrative multi-timescale control loop for ME-STAR-RIS-enabled NTNs. Exact update rates are hardware- and scenario-dependent; the key principle is to avoid moving elements at the same rate as electronic STAR reconfiguration.}
% \label{fig:control}
% \end{figure}

\subsection{Predictive and Learning-Assisted Configuration}
The deterministic nature of satellite ephemerides creates an opportunity for predictive control. Future user geometry and satellite elevation can often be anticipated over useful horizons, enabling the surface to reposition before the link quality deteriorates. Machine learning can complement model-based optimization by learning when movement is beneficial, providing warm starts for nonconvex solvers, or selecting among a library of geometry states. The most promising role of learning is therefore not to replace physical models, but to reduce online search and control overhead.

Digital twins are another natural fit. A twin can maintain the surface geometry, calibration state, actuator health, satellite trajectory, and environment model. It can evaluate candidate movements before execution and estimate whether a small gain justifies mechanical wear or energy expenditure.

%\begin{table}[t]
%\caption{Numerical Case-Study Parameters}
%\label{tab:case}
%\centering
%\footnotesize
%\begin{tabular}{ll}
%\toprule
%Parameter & Value/range \\
%\midrule
%LEO altitude & 550 km \\
%Carrier frequency / bandwidth & 2 GHz / 1 MHz \\ Satellite beam gain & 40 dBi \\ Effective surface-element / user gain & 6 dBi / 0 dBi \\ Surface--user distance & 200 m on each side \\ Rician factor / diffuse paths & 2 dB / 4 \\ Default STAR-RIS size & 64 elements ($8\times8$) \\ Element sweep & 16, 36, 64, 100, 144 \\ Nominal movement-cell spacing & $1.5\lambda$ \\ Movement range $D$ & 0--$1\lambda$ \\ STAR energy split & 0.5 transmission / 0.5 reflection \\ Satellite transmit-power sweep & 40--60 dBm \\ Satellite beamforming & Fixed beam \\ Direct satellite--user link & Blocked \\ Channel knowledge & Perfect cascaded CSI\\ Monte Carlo realizations & 200 \\ Compared schemes & ME-STAR-RIS; fixed STAR-RIS \\
%\bottomrule
%\end{tabular}
%\end{table}

\section{Numerical Case Study}

A compact numerical study is used to quantify the additional gain offered by geometry reconfiguration without turning the article into a full optimization paper. The comparison is deliberately restricted to two schemes: the proposed ME-STAR-RIS and an otherwise identical fixed-position STAR-RIS. Both schemes operate under the same satellite link budget, user locations, propagation realizations, STAR energy-splitting protocol, number of surface elements, transmit power, satellite beam, and electronic STAR control. Hence, the only additional design freedom available to the proposed scheme is the ability to reposition the STAR elements within their prescribed local movement regions. This controlled setup allows the performance difference to be attributed directly to element mobility rather than to different beamforming, power-allocation, or channel assumptions.

\subsection{System, Channel, and Optimization Setup}

\begin{figure*}[t]
\centering
\begin{minipage}[t]{0.45\textwidth}
\centering
\includegraphics[width=\linewidth]{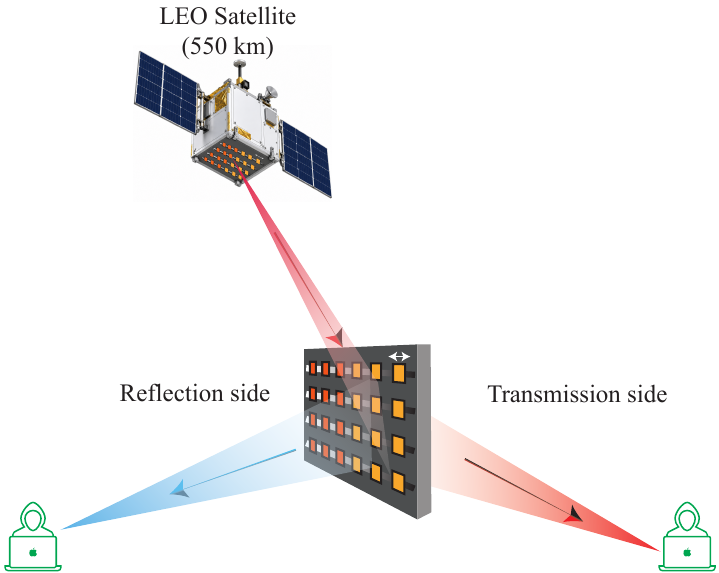}\\[-0.6ex]
{\footnotesize (a)}
\end{minipage}\hfill
\begin{minipage}[t]{0.45\textwidth}
\centering
\includegraphics[width=\linewidth]{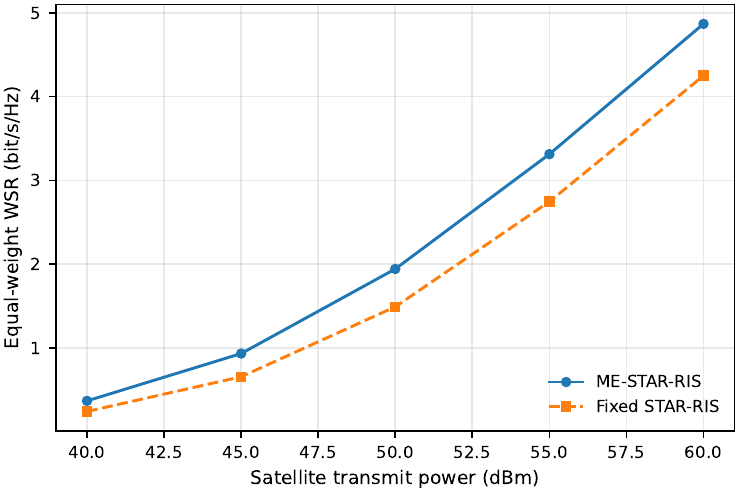}\\[-0.6ex]
{\footnotesize (b)}
\end{minipage}

\vspace{0.6ex}
\begin{minipage}[t]{0.45\textwidth}
\centering
\includegraphics[width=\linewidth]{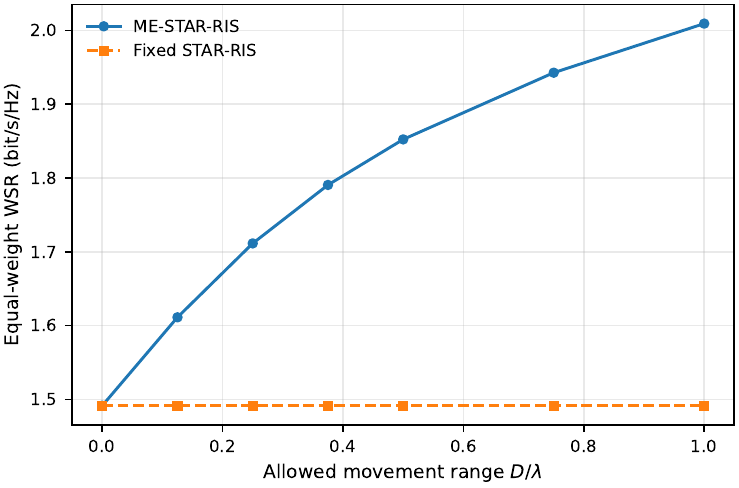}\\[-0.6ex]
{\footnotesize (c)}
\end{minipage}\hfill
\begin{minipage}[t]{0.45\textwidth}
\centering
\includegraphics[width=\linewidth]{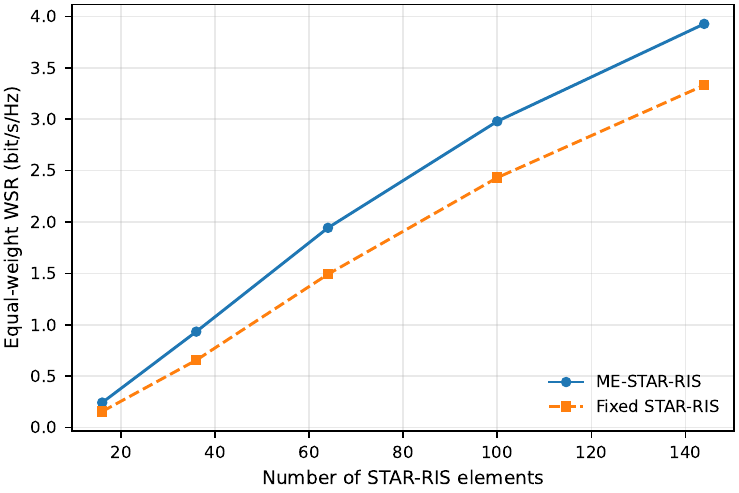}\\[-0.6ex]
{\footnotesize (d)}
\end{minipage}
\caption{LEO-centered numerical comparison between ME-STAR-RIS and an otherwise identical fixed STAR-RIS: (a) Gateway-mounted full-space STAR geometry considered in the case study, (b) Equal-weight WSR versus satellite transmit power for $N=64$ and $D=0.75\lambda$, (c) Equal-weight WSR versus allowable movement range for $N=64$ and 50~dBm, and (d) Equal-weight WSR versus the number of STAR elements for $D=0.75\lambda$ and 50~dBm. }
\label{fig:case}
\end{figure*}

We consider a LEO satellite at an altitude of 550~km illuminating a gateway-mounted STAR surface, with one ground user located on the reflection side and another on the transmission side, as illustrated in Fig.~\ref{fig:case}(a). The carrier frequency is 2~GHz and the bandwidth is 1~MHz. The satellite forms a high-gain beam toward the STAR surface with an antenna gain of 40~dBi. An effective per-element link-budget gain of 6~dBi is adopted for the STAR surface, while the user antenna gain is 0~dBi, and the surface-to-user distance is 200~m for both users. The receiver noise temperature and noise figure are 290~K and 5~dB, respectively. These S-band parameters are intentionally selected for a compact passive-surface example in which the satellite-to-surface hop remains physically meaningful without requiring an excessively large passive aperture.

The satellite-to-surface channel is modeled as a far-field line-of-sight (LoS) link. In contrast, the surface-to-user channels follow a position-dependent Rician field-response model with a Rician factor of 2~dB and four diffuse components. Thus, changing an element position modifies the phases and amplitudes associated with the local multipath field and, consequently, the effective cascaded satellite--surface--user channel. Doppler is assumed to be compensated at the satellite/receiver level so that the numerical experiment focuses specifically on the geometry-reconfiguration mechanism rather than on carrier-frequency tracking.

The direct satellite-to-user links are deliberately omitted. This represents a shadowed or locally blocked service condition in which the gateway-mounted STAR surface provides the dominant propagation path. More importantly, excluding the direct component allows the experiment to isolate the additional gain produced by the reconfigurable surface geometry without introducing direct/cascaded coherent-combining effects. Consequently, the reported numerical gains should be interpreted as \emph{surface-assisted geometry gains} rather than as universal end-to-end gains for every satellite-user configuration.

The default STAR surface contains 64 elements arranged on an $8\times8$ nominal grid. Adjacent local movement-cell centers are separated by $1.5\lambda$. Each element can move along a one-dimensional local track of total length $D$. Therefore, for $D\leq\lambda$, adjacent elements in the same row remain separated by at least $0.5\lambda$, which prevents element overlap and avoids unrealistically small spacing. The fixed STAR-RIS uses the centers of exactly the same movement cells and therefore has the same nominal aperture and element count.

To further isolate the effect of element mobility, the satellite beamforming is not jointly optimized with the surface configuration. Instead, the same fixed high-gain satellite beam is directed toward the STAR surface for both schemes. This choice is intentional: if satellite beamforming were also adapted independently in every realization, part of the observed performance change could arise from the active-array optimization rather than from the movable aperture itself. The present study therefore asks a more focused question: \emph{given the same satellite illumination and the same electronic STAR capability, how much additional gain is obtained by allowing the surface geometry to change?}

The two users occupy equal-bandwidth orthogonal resource blocks and receive equal transmit power, which removes inter-user interference and keeps the case study centered on the surface response. Accordingly, the reported WSR is normalized to the total system bandwidth, with each user's spectral-efficiency contribution weighted by its one-half bandwidth share. The STAR energy-splitting factors are fixed at 0.5 for transmission and 0.5 for reflection in both schemes. For every channel realization and every candidate surface geometry, the transmission and reflection phases are reconfigured to coherently combine the corresponding cascaded channel contributions. Hence, the fixed STAR-RIS is \emph{not} an unoptimized baseline: its electronic STAR phases are optimized under the same ideal independent-phase energy-splitting model as those of the ME-STAR-RIS. The proposed scheme differs only in that it can additionally optimize the element locations. Practical coupled transmission/reflection phase constraints are intentionally excluded from this illustrative experiment and are discussed separately in Section~VI-A.

The resulting design objective is to maximize the equal-weight WSR subject to the feasible element-movement regions and spacing constraints. Since the satellite beam, power allocation, and energy-splitting factors are fixed, the remaining controllable quantities are the electronic STAR phases for both schemes and, additionally, the element positions for ME-STAR-RIS. Perfect knowledge of the considered cascaded channel realization is assumed during this numerical optimization. This assumption provides a controlled benchmark for assessing the potential geometry gain; the additional difficulty of acquiring position-dependent CSI under LEO mobility is addressed in Section~IV-B.

For the ME-STAR-RIS, the position problem is handled using a low-complexity element-wise coordinate search. Each element is sequentially evaluated over 11 candidate locations on its local track, while the electronic STAR phases are realigned for the resulting candidate geometry. A candidate position is accepted only if it increases the equal-weight WSR. Two complete sweeps over all surface elements are performed. This replaces an exhaustive joint search over all element-position combinations with sequential one-dimensional searches. For a fixed number of candidate locations and sweeps, the number of position evaluations therefore scales linearly with the number of movable elements. The fixed STAR-RIS requires only the electronic phase configuration because its element locations remain at the cell centers. All reported results are averaged over 200 independent channel realizations. Identical realizations are used for the ME-STAR-RIS and fixed STAR-RIS, and the same 200-realization set is reused across the parameter sweeps. Consequently, the common operating point ($N=64$, $D=0.75\lambda$, and 50~dBm) is numerically identical in Figs.~\ref{fig:case}(b)--(d).

\subsection{Results and Design Insights}

Figure~\ref{fig:case}(b) evaluates the equal-weight WSR as the satellite transmit power increases. The ME-STAR-RIS consistently outperforms the fixed STAR-RIS over the entire considered range, showing that geometry reconfiguration provides a gain that is complementary to transmit-power scaling. At 50~dBm, the fixed STAR-RIS achieves 1.491~bit/s/Hz, whereas the ME-STAR-RIS reaches 1.943~bit/s/Hz, corresponding to a 30.3\% improvement. The relative gain is larger in the low-to-moderate-SNR regime, reaching 52.1\% at 40~dBm and 42.2\% at 45~dBm, but decreases to 14.5\% at 60~dBm.

This trend has a clear physical interpretation. Repositioning the elements creates a more favorable effective cascaded channel before the electronic STAR phases perform coherent combining. At low and moderate SNR, this additional channel gain translates directly into a substantial rate improvement. As the received SNR becomes large, however, the logarithmic rate function compresses the relative benefit of further effective-channel enhancement. Therefore, the percentage gain decreases with transmit power even though the ME-STAR-RIS remains uniformly superior in absolute WSR. The Monte Carlo standard errors are below 0.018~bit/s/Hz for all plotted mean values, indicating that the observed separation between the two schemes is well resolved using 200 realizations.

Figure~\ref{fig:case}(c) isolates the effect of the available movement range. When $D=0$, the ME-STAR-RIS reduces exactly to the fixed STAR-RIS and both schemes achieve approximately 1.491~bit/s/Hz. This coincidence provides an important internal consistency check because the proposed architecture should offer no additional degree of freedom when physical movement is disabled. Allowing only $0.25\lambda$ of local displacement increases the WSR to 1.711~bit/s/Hz, corresponding to a 14.7\% gain. The improvement rises to 24.2\% at $D=0.5\lambda$, 30.3\% at $D=0.75\lambda$, and 34.7\% at $D=\lambda$.

The curve also reveals diminishing returns as the allowable track length increases. A relatively small displacement already allows many elements to move toward locally favorable points of the spatial field. Once a sufficiently useful set of phases and path combinations has been reached, increasing the movement region further produces progressively smaller additional improvements. This observation is particularly relevant to NTN implementation: it suggests that large or continuously moving mechanical structures may not be necessary. A bounded local displacement, invoked only when the predicted geometry gain justifies the actuation cost, may capture a substantial fraction of the available benefit. This result therefore supports the multi-timescale architecture discussed in Section~IV-C, in which physical geometry is updated relatively slowly while electronic STAR coefficients provide fast adaptation.

Figure~\ref{fig:case}(d) examines the effect of the number of STAR elements. As expected, increasing the surface size improves the WSR of both architectures because a larger aperture provides stronger coherent combining. Nevertheless, ME-STAR-RIS remains superior for every considered surface size. At the default 64-element configuration, the mobility gain is 30.3\%. With 144 elements, the WSR increases from 3.330~bit/s/Hz for the fixed STAR-RIS to 3.927~bit/s/Hz for ME-STAR-RIS, corresponding to an absolute improvement of approximately 0.597~bit/s/Hz.

An interesting trend is that the \emph{relative} mobility gain decreases as the number of elements becomes large, even though the absolute rate difference remains significant. A large fixed STAR aperture already offers substantial array and coherent-combining gain, which reduces the relative importance of the additional geometry degree of freedom. Conversely, when the number of surface elements is limited by payload size, mass, cost, or power consumption, geometry optimization becomes particularly valuable because it extracts more performance from a constrained aperture.

Taken together, Figs.~\ref{fig:case}(b)--(d) reveal three complementary design messages. Increasing transmit power improves both systems but does not eliminate the mobility advantage; enlarging the permitted movement region increases the geometry gain but with diminishing returns; and increasing the number of surface elements reduces the relative benefit of movement because the fixed aperture itself becomes increasingly capable. Thus, ME-STAR-RIS should not be viewed simply as an alternative way of adding transmit power or surface elements. Instead, element mobility provides an additional geometry-control degree of freedom that is particularly valuable when power, aperture size, or deployment flexibility is constrained--precisely the operating conditions that frequently arise in non-terrestrial networks.

\section{Challenges and Future Research Directions}
While ME-STAR-RIS introduces an additional geometry degree of freedom
for NTN design, translating this flexibility into practical deployments
requires addressing several hardware, channel-acquisition, control, and
reliability issues. At the same time, geometry reconfiguration creates
new opportunities for near-field operation, sensing and security, and
cooperative multi-platform networking. This section summarizes the
main remaining challenges and the research directions that can enable
practical ME-STAR-RIS-assisted NTNs.

\subsection{Hardware and Electromagnetic Constraints}

Physical movement introduces actuation energy, repositioning
latency, mechanical wear, and calibration requirements that are
absent from conventional fixed STAR-RISs. These issues are
particularly important for satellite-borne implementations, where
repeated movement may induce micro-vibration and interact with
thermal and attitude-control constraints. Hence, practical designs
are likely to favor limited-stroke, group-wise, or event-triggered
movement rather than continuous element repositioning.

Movement also changes inter-element spacing, mutual coupling,
radiation characteristics, and the transmission/reflection response.
This is especially important because practical STAR-RIS hardware
may already impose coupling between the transmission and reflection
coefficients \cite{liu2022coupled}. Future models should therefore
jointly capture element position, orientation, coupling, amplitude,
phase, loss, finite-resolution actuation, and position uncertainty.
Hardware prototypes and space-relevant measurements are ultimately
needed to establish how much of the ideal geometry gain can be
retained in practice.

\subsection{CSI and Scalable Control}

ME-STAR-RIS requires knowledge of both the electromagnetic state
and the physical geometry of the surface. Estimating the channel
independently for every possible element configuration is impractical,
particularly in rapidly varying NTN environments. Statistical CSI,
satellite ephemerides, geometric channel models, compressed
measurements, spatial channel maps, and position sensing provide
promising alternatives. Group-wise movement and geometry codebooks
can further reduce signaling overhead.

Scalability is another major challenge. Adding element coordinates
to the already coupled STAR coefficient, beamforming, power, and
scheduling variables produces a highly nonconvex problem. Existing
ME-STARS works demonstrate alternating-optimization approaches in
terrestrial systems \cite{zhao2026mestars,zhao2026secure}, but
network-scale NTN implementations will require decomposition,
distributed optimization, and learning-assisted initialization.
Rather than replacing physical models, prediction and learning may
be particularly useful for deciding \emph{when} geometry
reconfiguration is worthwhile.

\subsection{Near-Field, Sensing, and Security}

Electrically large ME-STAR-RIS deployments at gateways, HAPS, or
satellite platforms may operate partly in the near field, where
element position affects both angle and range focusing. Geometry
reconfiguration could therefore support three-dimensional user
separation, localized energy delivery, and adaptive focusing, while
also creating new calibration and channel-estimation requirements.

The same spatial flexibility is attractive for sensing, localization,
and security. Movable-element STAR-RIS has already been considered
for integrated sensing and covert communication in terrestrial
networks \cite{zhou2025iscc}. In NTNs, element movement could reshape
sensing beampatterns, improve localization geometry, suppress clutter,
or create spatial nulls toward eavesdroppers and jammers. However,
the geometry-control link itself becomes a potential attack surface,
which motivates authenticated movement commands, robust geometry
feedback, and secure fallback configurations.

\subsection{Networks Cooperation, Reliability, and Standardization}

Future ME-STAR-RIS systems need not operate as isolated surfaces.
Multiple satellites, HAPS, gateways, and STAR surfaces could
coordinate their geometry and electronic states, with only the nodes
that provide sufficient system-level benefit being physically
reconfigured. Such sparse movement could retain much of the
geometry gain while reducing total actuation energy and control
overhead.

Reliability must also become part of the optimization. Stuck
elements, actuator saturation, position drift, calibration loss, and
control-link failures should not interrupt service; a natural fallback
is to operate affected elements as a fixed STAR-RIS. Digital twins
can track actuator health, accumulated movement, calibration, and
thermal state so that geometry updates are performed only when their
expected benefit justifies the associated wear. Finally, practical
deployment will require standard mechanisms for geometry-state
signaling, calibration, failure reporting, and coordination with NTN
beam management and handover procedures.

\section{Conclusion}

ME-STAR-RIS extends the programmable-surface paradigm from purely
electromagnetic reconfiguration to joint electromagnetic and geometric
adaptation. This additional degree of freedom is particularly attractive
for 6G NTNs, where satellite, aerial, and terrestrial links are strongly
influenced by time-varying propagation geometry. We presented
representative satellite-borne, aerial-relay, and cross-tier architectures
and discussed the enabling mechanisms required for their operation,
including geometry-aware optimization, channel acquisition,
multi-timescale control, and predictive configuration. The numerical case study showed that movable-element operation can consistently outperform an otherwise identical fixed STAR-RIS, while also revealing important design tradeoffs. In particular, the benefit of element mobility exhibits diminishing returns as the allowable movement range increases and becomes relatively smaller as the surface aperture grows. These observations reinforce a central design principle:
physical element repositioning should be used selectively and at a
slower or event-triggered timescale, while electronic STAR
reconfiguration handles faster channel variations. Moving toward
practical ME-STAR-RIS-enabled NTNs will therefore require
hardware-aware electromagnetic models, efficient geometry control,
reliable actuation and calibration, and experimental validation of the
performance--complexity tradeoff.

\bibliographystyle{IEEEtran}
\bibliography{references}

@ARTICLE{11674230,
  author={Khan, Abd Ullah and Khan, Wali Ullah and Jung, Haejoon and Shin, Hyundong},
  journal={IEEE Trans. Commun.}, 
  title={Resource Allocation and Sharing for {UAV-assisted integrated TN-NTN} with Multi-Connectivity}, 
  year={2026},
  volume={},
  number={},
  pages={1-1},
  doi={10.1109/TCOMM.2026.3729734}}

@ARTICLE{10716670,
  author={Khan, Wali Ullah and Mahmood, Asad and others},
  journal={IEEE Network}, 
  title={Beyond Diagonal {RIS for 6G} Non-Terrestrial Networks: {Potentials} and Challenges}, 
  year={2025},
  volume={39},
  number={1},
  pages={80-89},
  doi={10.1109/MNET.2024.3480332}}

@ARTICLE{10133841,
  author={Ahmed, Manzoor and Wahid, Abdul and others},
  journal={IEEE Internet of Things J.}, 
  title={A Survey on {STAR-RIS: Use} Cases, Recent Advances, and Future Research Challenges}, 
  year={2023},
  volume={10},
  number={16},
  pages={14689-14711},
  doi={10.1109/JIOT.2023.3279357}}

@article{liu2022coupled,
  author  = {Y. Liu and X. Mu and others},
  title   = {Simultaneously Transmitting and Reflecting ({STAR})-{RIS}s: A Coupled Phase-Shift Model},
  journal = {IEEE Wireless Commun.},
  volume  = {29},
  number  = {2},
  pages   = {169--176},
  year    = {2022},
  doi     = {10.1109/MWC.102.2100363}
}

@article{zhu2024ma,
  author  = {L. Zhu and W. Ma and R. Zhang},
  title   = {Movable Antennas for Wireless Communication: Opportunities and Challenges},
  journal = {IEEE Commun. Mag.},
  volume  = {62},
  number  = {6},
  pages   = {114--120},
  year    = {2024},
  doi     = {10.1109/MCOM.001.2300212}
}

@article{zhu2025tutorial,
  author  = {L. Zhu and W. Ma and others},
  title   = {A Tutorial on Movable Antennas for Wireless Networks},
  journal = {IEEE Commun. Surveys Tuts.},
  volume  = {28},
  pages   = {3002--3054},
  year    = {2026},
  doi     = {10.1109/COMST.2025.3546373}
}

@article{zhou2026meris,
  author  = {D. Zhou and W. Mei and others},
  title   = {Movable-Element {RIS}: Joint Element Positioning and Beamforming Optimization},
  journal = {IEEE Wireless Commun. Lett.},
  volume  = {15},
  pages   = {915--919},
  year    = {2026},
  doi     = {10.1109/LWC.2025.3641191}
}

@article{zhao2026meris,
  author  = {J. Zhao and Q. Huang and others},
  title   = {Movable-Element {RIS}-Aided Wireless Communications: An Element-Wise Position Optimization Approach},
  journal = {IEEE Commun. Lett.},
  volume  = {30},
  pages   = {967--971},
  year    = {2026},
  doi     = {10.1109/LCOMM.2026.3655084}
}

@article{zhao2026mestars,
  author  = {J. Zhao and Q. Zhou and others},
  title   = {Exploiting Movable-Element {STARS} for Wireless Communications},
  journal = {IEEE Trans. Wireless Commun.},
  volume  = {25},
  pages   = {11465--11479},
  year    = {2026},
  doi     = {10.1109/TWC.2026.3659604}
}

@article{zhao2026secure,
  author  = {J. Zhao and Q. Xu and others},
  title   = {Movable-Element {STARS}-Aided Secure Communications},
  journal = {IEEE Trans. Veh. Technol.},
  volume  = {75},
  number  = {3},
  pages   = {4608--4621},
  year    = {2026},
  doi     = {10.1109/TVT.2025.3615115}
}

@article{zhou2025iscc,
  author  = {T. Zhou and K. Xu and others},
  title   = {{STAR-RIS}-Empowered Integrated Sensing and Covert Communication System With Movable Elements: Joint Robust Beamforming and Element Deployment Design},
  journal = {IEEE Trans. Cogn. Commun. Netw.},
  volume  = {11},
  number  = {5},
  pages   = {2893--2909},
  year    = {2025},
  doi     = {10.1109/TCCN.2025.3588724}
}

@article{amiri2025swipt,
  author  = {M. Amiri and A. Mohammadzadeh and others},
  title   = {Movable Antenna {SWIPT} Systems With {STAR-RIS}: A Meta-{DRL} Approach},
  journal = {IEEE Trans. Veh. Technol.},
  volume  = {75},
  number  = {4},
  pages   = {6864--6869},
  year    = {2026},
  doi     = {10.1109/TVT.2025.3622305}
}

@article{lin2024simleo,
  author  = {S. Lin and J. An and others},
  title   = {Stacked Intelligent Metasurface Enabled {LEO} Satellite Communications Relying on Statistical {CSI}},
  journal = {IEEE Wireless Commun. Lett.},
  volume  = {13},
  number  = {5},
  pages   = {1295--1299},
  year    = {2024},
  doi     = {10.1109/LWC.2024.3368238}
}

@article{zhao2022staruav,
  author  = {J. Zhao and Y. Zhu and others},
  title   = {Simultaneously Transmitting and Reflecting Reconfigurable Intelligent Surface ({STAR-RIS}) Assisted {UAV} Communications},
  journal = {IEEE J. Sel. Areas Commun.},
  volume  = {40},
  number  = {10},
  pages   = {3041--3056},
  year    = {2022},
  doi     = {10.1109/JSAC.2022.3196102}
}

@article{zheng2026satris,
  author  = {Z. Zheng and X. Li and others},
  title   = {Reconfigurable Intelligent Surface-Enhanced Satellite Networks: Deployment Strategies, Key Capabilities, Practical Solutions, and Future Directions},
  journal = {npj Wireless Technol.},
  volume  = {2},
  pages   = {44},
  year    = {2026},
  doi     = {10.1038/s44459-026-00051-0}
}

\end{document}